\documentclass[usenatbib]{mn2e}
\usepackage{epsfig}

\newif\ifAMStwofonts
\AMStwofontstrue

\def\be{\begin{equation}}
\def\ee{\end{equation}}

\def\gtsima{$\; \buildrel > \over \sim \;$}
\def\ltsima{$\; \buildrel < \over \sim \;$}
\def\prosima{$\; \buildrel \propto \over \sim \;$}
\def\gsim{\lower.5ex\hbox{\gtsima}}
\def\lsim{\lower.5ex\hbox{\ltsima}}
\def\simgt{\lower.5ex\hbox{\gtsima}}
\def\simlt{\lower.5ex\hbox{\ltsima}}
\def\simpr{\lower.5ex\hbox{\prosima}}

\def\HI{\hbox{H$\,\rm \scriptstyle I\ $}}
\def\HII{\hbox{H$\,\rm \scriptstyle II\ $}}

\def\HeII{\hbox{He$\,\rm \scriptstyle II\ $}}

\title[The 21~cm forest with LOFAR]{Prospects for detecting the 21~cm forest from the diffuse intergalactic medium with LOFAR}

\author[B. Ciardi et al.]
{B. Ciardi$^1$\thanks{E-mail:ciardi@mpa-garching.mpg.de}, P. Labropoulos$^2$,
 A. Maselli$^3$, R. Thomas$^4$, S. Zaroubi$^5$, L. Graziani$^1$, \and
 J.~S. Bolton$^6$, G. Bernardi$^7$, M. Brentjens$^2$, A.~G. de Bruyn$^{5,2}$, S. Daiboo$^5$, \and
 G.~J.~A. Harker$^8$, V. Jelic$^2$, S. Kazemi$^5$, L.~V.~E. Koopmans$^5$, O. Martinez$^5$, \and
 G. Mellema$^9$,  A.~R. Offringa$^5$, V.~N. Pandey$^{5,2}$, J. Schaye$^{10}$, V. Veligatla$^5$, \and
 H. Vedantham$^5$, S. Yatawatta$^{4,2}$
 \\
$^1$ Max-Planck-Institut fuer Astrophysik,
     Karl-Schwarzschild-Strasse 1, D-85748 Garching b. Muenchen, Germany\\
$^2$ ASTRON, PO Box 2, 7990 AA Dwingeloo, the Netherlands\\
$^3$ EVENT Lab for Neuroscience and Technology, Universitat de Barcelona, 
     Passeig de la Vall d'Hebron 171, 08035 Barcelona, Spain\\
$^4$ CITA, University of Toronto, 60 St George Street, M5S 3H8, Toronto, ON, Canada\\
$^5$ Kapteyn Astronomical Institute, University of Groningen, PO Box 800, 9700 AV Groningen, the Netherlands\\
$^6$ School of Physics, University of Melbourne, Parkville, Victoria 3010, Australia\\
$^7$ Harvard Smithsonian Center for Astrophysics, 60 Garden Street, Cambridge, MA 02138, USA\\
$^8$ Center for Astrophysics and Space Astronomy, University of Colorado Boulder, CO 80309, USA\\
$^9$ Department of Astronomy and Oskar Klein Centre for Cosmoparticle Physics, AlbaNova, Stockholm University, SE-106 91 Stockholm, Sweden\\
$^{10}$ Leiden Observatory, Leiden University, PO Box 9513, 2300RA Leiden, the Netherlands\\
}

\date{October 2012}          

\pagerange{\pageref{firstpage}--\pageref{lastpage}}
\pubyear{2012}
\begin{document}

\maketitle
\label{firstpage}

\begin{abstract}
  We discuss the feasibility of the detection of the 21~cm forest in
  the diffuse IGM with the radio telescope {\tt LOFAR}. The optical
  depth to the 21~cm line has been derived using simulations of
  reionization which include detailed radiative transfer of ionizing
  photons. We find that the spectra from reionization models with
  similar total comoving hydrogen ionizing emissivity but different
  frequency distribution look remarkably similar.  Thus, unless the
  reionization histories are very different from each other (e.g. a
  predominance of UV vs. x-ray heating) we do not expect to
  distinguish them by means of observations of the 21~cm forest.
  Because the presence of a strong x-ray background would make the
  detection of 21~cm line absorption impossible, the lack of
  absorption could be used as a probe of the presence/intensity of the
  x-ray background and the thermal history of the universe.  Along a
  random line of sight {\tt LOFAR} could detect a global suppression
  of the spectrum from $z$ \gsim 12, when the IGM is still mostly
  neutral and cold, in contrast with the more well-defined, albeit broad,
  absorption features visible at lower redshift. Sharp, strong
  absorption features associated with rare, high density pockets of
  gas could be detected also at $z \sim 7$ along preferential lines of
  sight.
\end{abstract}



\section{Introduction}         

The hyperfine transition of the ground state of the neutral hydrogen
atom, $^2S_{1/2}$, is the most direct probe of the content and
distribution of neutral hydrogen in the universe.  Recent advances in
the construction and planning of powerful radio interferometers
(e.g. {\tt LOFAR}\footnote{http://lofar.org}, {\tt
MWA}\footnote{http://www.mwatelescope.org}, {\tt
SKA}\footnote{http://www.skatelescope.org}) are promising the
opportunity to make direct observations of the high redshift
intergalactic medium (IGM), at $z \gsim 6.5$.

The attention dedicated to 21~cm studies in cosmology so far has been
mainly focused on 21~cm tomography, which would ideally provide a 3D
mapping of the evolution of neutral hydrogen
(e.g. \citealt{Madau.Meiksin.Rees_1997,Tozzi.Madau.Meiksin.Rees_2000,
Ciardi.Madau_2003,Mellema.Iliev.Pen.Shapiro_2006,Datta_etal_2008,Lidz_etal_2008b,Santos_etal_2008,
Baek_etal_2009,Geil.Wyithe_2009,Morales.Wyithe_2010,Santos_etal_2010}).
Although this is an extremely exciting prospect, it suffers from
several severe difficulties
(e.g. \citealt{Shaver.Windhorst.Madau.deBruyn_1999,
DiMatteo.Perna.Abel.Rees_2002,
DiMatteo.Ciardi.Miniati_2004,Gleser.Nusser.Benson_2008,Geil.Wyithe.Petrovic.Oh_2008,
Jelic_etal_2008,Bernardi_etal_2009,Bowman.Morales.Hewitt_2009,Liu_etal_2009,
Bernardi_etal_2010}), the most relevant being foreground and
ionospheric contamination, terrestrial interference, low spatial
resolution and the requirement that the IGM spin temperature, $T_{S}$,
is decoupled from the temperature of the cosmic microwave background
(CMB), $T_{\rm CMB}$.

A valid alternative to tomography is looking at the 21~cm lines
generated in absorption against high-$z$ radio loud sources by the
neutral IGM and intervening collapsed structures, i.e. to search for
the 21~cm forest (e.g. \citealt{Carilli.Gnedin.Owen_2002,
Furlanetto.Loeb_2002,Furlanetto_2006,Carilli_etal_2007,Xu_etal_2009,Mack.Wyithe_2011,
Meiksin_2011,Xu.Ferrara.Chen_2011}). Analogous to the extensively
studied case of the Ly$\alpha$ forest (for a review see
\citealt{Meiksin_2009}), the 21~cm forest signal can be detected in
the spectra of high-$z$ radio sources and results from the absorption
produced by the intervening neutral hydrogen along the line of sight
(LOS), which removes from the spectrum the continuous radiation
locally redshifted to the resonance frequency.

Despite the great challenge posed by the requirement of sufficiently
bright, and hence extremely rare, target sources
(e.g. \citealt{Carilli.Gnedin.Owen_2002,Xu_etal_2009,Mack.Wyithe_2011}),
the 21~cm forest is particularly appealing as it naturally bypasses
the main limitations expected for 21~cm tomographic
measurements. Using 21~cm forest observations provides several
benefits: {\it (i)} the IGM is visible even if $T_s = T_{\rm CMB}$,
because the CMB does not act as background source; {\it (ii)} the IGM
can be spatially resolved on small scales, of the order of a few tens
of kpc or even lower; {\it (iii)} spectroscopy of bright point sources
is technically easier than 2D image processing, and all problems
associated with foregrounds and interference vanish \citep[see][for a
more detailed discussion]{Furlanetto_2006}.  A further advantage is
the close similarity with the Ly$\alpha$ forest, from which one can
adopt the well established and robust analysis techniques described in
the literature.

In this paper we discuss the possibility of detecting the 21~cm signal
with the radio telescope {\tt LOFAR}, by making use of the simulations
of reionization presented in
\citet[][C2012]{Ciardi.Bolton.Maselli.Graziani_2012}.  Differently
from previous studies mentioned above, the simulations used here are
run using a combination of hydrodynamic simulations and the detailed
radiative transfer of ionizing photons through a gas composed of
hydrogen and helium. The latter component is particularly important
for a correct determination of the gas temperature. In addition, the
expected observed spectrum is calculated based on the actual {\tt
LOFAR} configuration.  In Section~2 we briefly review the basic
physics relevant for the 21~cm forest signal. In Section~3 we discuss
the adopted simulations, while in Section~4 the method for producing
mock spectra of high-$z$ radio loud sources, as well as the dependence
on a number of quantities, is presented. In Section~5 we discuss the
observational feasibility and in Section~6 we give our conclusions.


\section{The 21cm forest}
\label{sec:basic}

As the physics behind the emission/absorption of the 21~cm line
has been extensively discussed in the literature by several authors
(for a review see e.g. \citealt*{Furlanetto.Oh.Briggs_2006}),
here we just write the relevant equations without derivation.
The 21~cm line corresponds to the hyperfine, spin-flip transition
of the ground state of the neutral hydrogen atom, $^2S_{1/2}$,
and to a rest frame frequency $\nu_{\rm 21cm}$ = 1.42~GHz.
A photon traveling through
a patch of neutral hydrogen at redshift $z$ will see the optical
depth \cite[e.g.][]{Madau.Meiksin.Rees_1997,Furlanetto.Oh.Briggs_2006}:

\begin{eqnarray}
\label{eq:tau}
\tau_{\rm 21cm}(z) & = & \frac{3}{32 \pi} \frac{h_p c^3 A_{\rm 21cm}}{k_B \nu_{\rm 21cm}^2}
\frac{x_{\rm HI} n_{\rm H}}{T_s (1+z) (dv_\parallel/dr_\parallel)}  \\ \nonumber
 & = & 9.6 \times 10^{-3} x_{\rm HI} (1+\delta) \left( \frac{1+z}{10} \right)^{3/2} \\ \nonumber
 &   & \left( \frac{T_{\rm CMB(z)}}{T_s(z)} \right) \left[ \frac{H(z)/(1+z)}{dv_\parallel/dr_\parallel} \right] ,
\end{eqnarray} 
where $n_{\rm H}$ is the H number density, $\delta$ is the gas density contrast,
$x_{\rm HI}$ is the mean neutral hydrogen fraction, $T_{\rm CMB}$ is the CMB temperature,
$T_s$ is the gas spin temperature (which quantifies the relative population
of the two levels of the $^2S_{1/2}$ transition), $A_{\rm 21cm}=2.85 \times 10^{-15}$~s$^{-1}$
is the Einstein coefficient of the transition, $H$ is the Hubble parameter and 
$dv_\parallel/dr_\parallel$ is the gradient of the proper velocity along the LOS 
(in km~s$^{-1}$), which takes into account also the contribution of the gas peculiar velocity. 
The other symbols appearing in the equation above have the standard meaning adopted in the
literature. 
It should be noted that here we do not perform a convolution with a Doppler line
profile. Although we do not expect this approximation to affect our estimates of the detectability of 
absorption lines with {\tt LOFAR}, we are planning to investigate this issue in more
detail in the future.

The photons emitted by a radio loud source at
redshift $z_s$ with frequencies $\nu>\nu_{\rm 21cm}$, will be removed from
the source spectrum with a probability $(1-e^{-\tau_{\rm 21cm}})$,
absorbed by the neutral hydrogen present along the LOS at
redshift $z=(\nu_{\rm 21cm}/\nu)(1 + z_s)-1$. Analogously to the case of
the Ly$\alpha$ forest, this could result in an average suppression
of the source flux (produced by diffuse neutral hydrogen),
as well as in a series of isolated absorption lines (produced by
overdense clumps of neutral hydrogen).

Due to the several orders of magnitude difference in the corresponding
Einstein coefficients, the physical regimes in which the 21~cm forest
and the Ly$\alpha$ forest are relevant, are significantly different.
As a consequence of the extremely low value of $A_{\rm 21cm}$, the
21~cm absorption is not saturated at high redshifts, as is the case
for the Ly$\alpha$ forest. This makes the 21~cm forest signal suitable
for studying the IGM in the epoch prior to reionization,
complementarily to the Ly$\alpha$ forest which instead is more
appropriate for post-reionization studies.  Nevertheless, the 21~cm
forest detection is in practice extremely hard, mainly due to the
tremendously weak 21~cm magnetic hyperfine transition. As can be
easily inferred from eq.~\ref{eq:tau}, the associated optical depth is
above 0.1, and could therefore produce an observable signal, only under
severe physical conditions (i.e. highly neutral, dense and cold gas).
Considering furthermore the paucity of radio loud sources at the high
redshifts of interest, and their anti-correlation with the optical
depth due to the fact that these kinds of objects are thought to be
associated with large ionizing fluxes, it is immediately clear that
the detection of a 21~cm forest signal is an extremely challenging
experiment.


\section{Simulations of reionization}
\label{sec:simul}

The modelling of mock spectra has been done in two
main steps. As can be seen from eq.~\ref{eq:tau}, the optical depth and
thus the absorption along the line of sight to a high redshift
source, depends significantly on the gas spin temperature,
as well as on the neutral hydrogen fraction at each point
along the LOS. The evolution of both these quantities
and their spatial fluctuations are determined by the
complex process of reionization, which needs to be properly
modelled. The first step in our spectra modelling pipeline, which is
described below, is then to simulate realistic reionization scenarios.
The second step is to use the outputs
of such simulations to extract random LOS and build mock absorption spectra of sources at
different redshifts. More details are given below in Section~\ref{sec:spectra}.
Unless stated otherwise, we assume that the spin temperature is coupled to the
gas temperature, $T$.

The simulations we use are described in C2012. Here we just
give some basic information and we refer the reader to the original paper
for more details.

\begin{figure}
\centering
\centering
\includegraphics[width=0.25\textwidth]{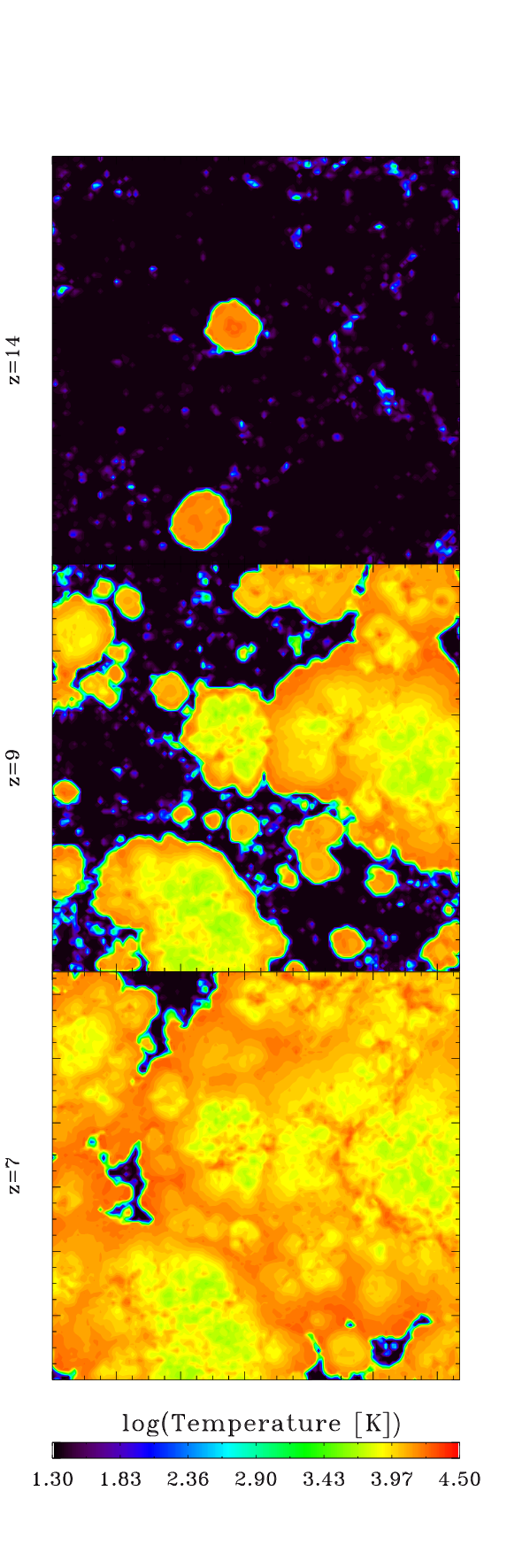}
\hspace*{-0.25in}
\includegraphics[width=0.25\textwidth]{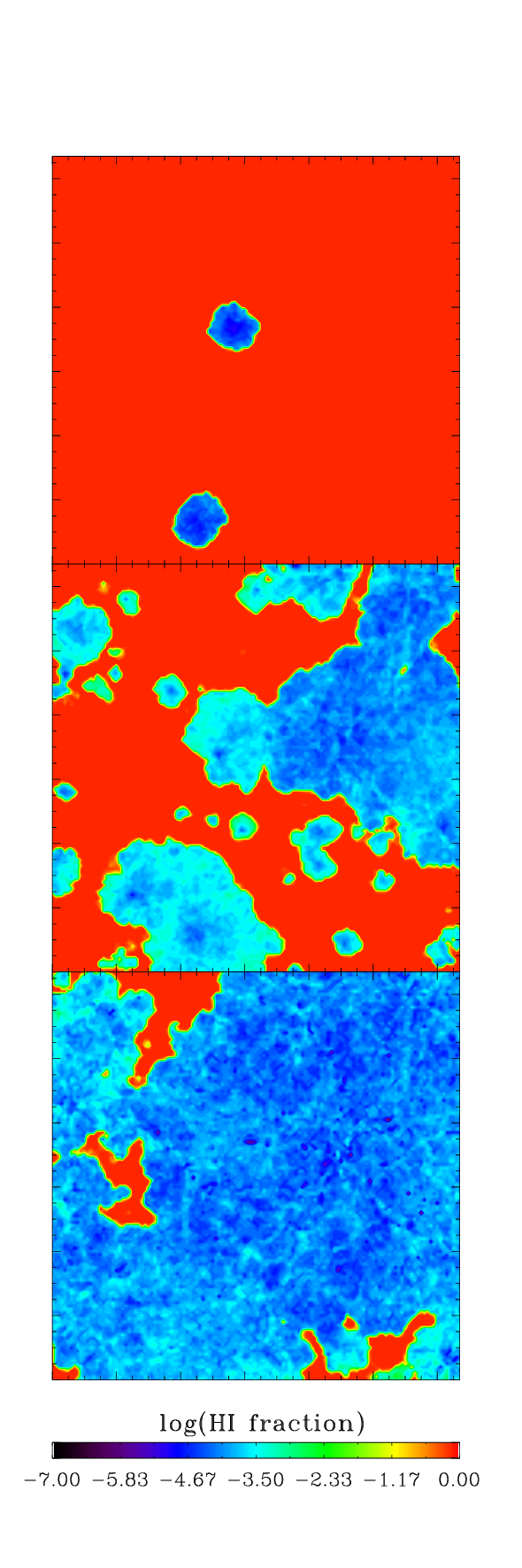}
\caption{{\it Left panel:} maps of temperature at redshift $z=14$ (upper panel), 9 (central
panel) and 7 (lower panel) for the simulation ${\mathcal E}$1.2-$\alpha$3. 
Each map represents the central slice of the simulation box.
{\it Right panel:} as the left panel but for the HI fraction.
}
\label{fig:maps}
\end{figure}

The simulations are based on a combination of hydrodynamic simulations
performed using the parallel Tree-SPH code {\small GADGET-3}
\citep[which is an updated version of the publicly available code
{\small GADGET-2}; see][]{Springel_2005} and of 3D radiative transfer
of both H and He followed by the Monte Carlo code {\tt CRASH}
(\citealt{Ciardi_etal_2001,Maselli.Ferrara.Ciardi_2003,Maselli.Ciardi.Kanekar_2009,
Pierleoni.Maselli.Ciardi_2009,Partl_etal_2011}; Graziani et al. in
prep.).  The hydrodynamic simulations were run in boxes of comoving
size 35.12$h^{-1}$~Mpc, with cosmological parameters
$\Omega_{\Lambda}=0.74$, $\Omega_m=0.26$, $\Omega_b=0.024 h^{-2}$,
$h=0.72$, $n_s=0.95$ and $\sigma_8=0.85$, where the symbols have the
usual meanings.  A total of $2 \times 512^3$ dark matter and gas
particles were followed in the simulation, yielding a mass per gas
particle of $4.15 \times 10^6 h^{-1}$~M$_\odot$.  After the simulation
outputs were obtained, the gas densities, peculiar velocities,
temperatures and halo masses were interpolated on to a uniform
128$^3$ grid for the radiative transfer calculations. The emissivity
and spectra of the sources were chosen based on a semi-analytic model
which satisfies a number of observational constraints, among them the
Thomson scattering optical depth and the \HI photoionization rate
measured from the Ly$\alpha$ forest.  The spectra adopted are always
power-laws, with index $\alpha$=1, 1.8 or 3. Our reference run
(${\mathcal E}$1.2-$\alpha$3 in the original paper) has $\alpha=3$. In
Figure~\ref{fig:maps} maps of the temperature (left panels) and
neutral fraction (right panels) are shown at redshifts $z=14, 9$ and
7, which correspond to volume-averaged \HII fractions $x_{\rm HII}$
(gas temperature $T$) of 0.029, 0.481 and 0.852 (488, 6020 and
10643~K), respectively. This simulation mimics a late reionization
model, while the other three simulations have been run with a higher
comoving hydrogen ionizing emissivity (see eq.~3 of C2012) and reach
full reionization at an earlier time. These all have a similar
emissivity assuming that 30\% (70\%) of the sources have $\alpha=1$
(3) (model ${\mathcal E}$1.2-$\alpha$1-3), or that all sources have
$\alpha=3$ (model ${\mathcal E}$1.6-$\alpha$3) or $\alpha=1.8$ (model
${\mathcal E}$1.2-$\alpha$1.8).  The characteristics of the models are
summarized in Table~1 of C2012. As mentioned, with the exception of
${\mathcal E}$1.2-$\alpha$3 all the simulations have similar comoving
hydrogen ionizing emissivities, so any differences in the reionization
histories are only due to the different spectral energy
distributions. The evolution of $x_{\rm HII}$ is very similar in all
models by construction, while \HeII reionization is completed
progressively later in models ${\mathcal E}$1.6-$\alpha$3 and
${\mathcal E}$1.2-$\alpha$3, which have softer ionizing spectra.  For
more details on the results we refer the reader to C2012.

\begin{figure}
\centering
\includegraphics[width=0.40\textwidth]{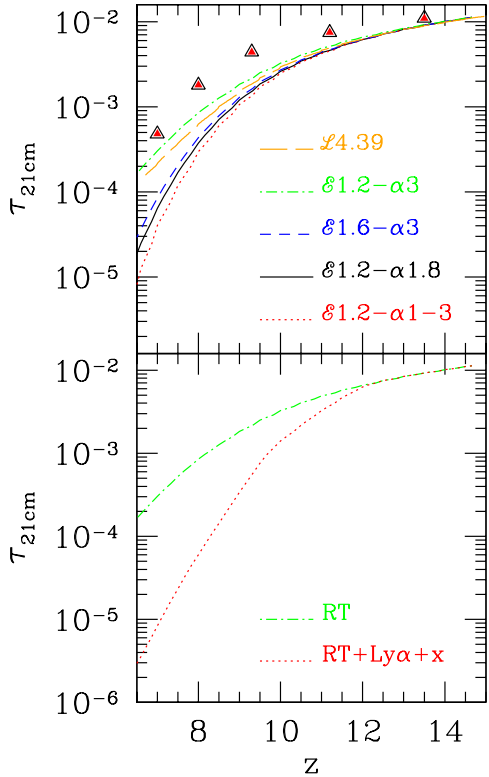}
\caption{{\it Upper panel}: evolution of the volume-averaged 21~cm
  optical depth for model ${\mathcal E}$1.2-$\alpha$1.8 (black solid
  line), ${\mathcal E}$1.2-$\alpha$1-3 (red dotted), ${\mathcal
  E}$1.6-$\alpha$3 (blue dashed), ${\mathcal E}$1.2-$\alpha$3 (green
  dotted-dashed) and ${\mathcal L}$4.39 (orange
  long-dashed). Triangles refer to the values in the simulation by
  \citet{Xu_etal_2009}.  {\it Bottom panel}: evolution of the volume
  averaged 21~cm optical depth in model ${\mathcal E}$1.2-$\alpha$3
  for different assumptions about the spin temperature, $T_s$, being:
  the same as the gas temperature, $T$, determined from the radiative
  transfer simulation (green dotted-dashed line); the maximum between
  $T$ and the temperature calculated including also the contribution
  from Ly$\alpha$ photons from stars and x-ray photons from quasars
  (red dotted line).  See text for more details.  }
\label{fig:tau21cm}
\end{figure}

It should be noticed that the outputs of the simulations are
sufficiently close to each other in redshift space that there is
always some overlap in the redshift range spanned by subsequent
boxes. In particular, we will be using boxes $B_i$ (with $i=1...54$)
which span the redshift range $z \approx 6-15$.

While the above simulations are useful to investigate the effect of
different reionization histories on the observability of the 21~cm
absorption features, we have run an additional, higher resolution
simulation in a 4.39$h^{-1}$~Mpc box with the aim of exploring instead
the impact of resolution on our results.  This simulation has been run
with $2 \times 256^3$ gas and dark matter particles, corresponding to
a mass of the gas particle of $6.48 \times 10^4 h^{-1}$~M$_\odot$, and
thus it is better suited to capture the absorption features due to
high density gas at smaller scales.  The simulation is similar to the
one analysed and discussed in Jeeson Daniel et al. (in prep.), but
here we have chosen an amplitude of the total comoving hydrogen
ionizing emissivity ${\mathcal E}=1.8$ (see eq.~3 in C2012) and
$\alpha=3$ (in the remaining we will refer to this simulation as
${\mathcal L}$4.39).  These choices were motivated by the requirement
that the evolution of the neutral fraction was similar to that of our
reference run ${\mathcal E}$1.2-$\alpha$3. As a reference, simulation
${\mathcal E}$1.2-$\alpha$3 (${\mathcal L}$4.39) has $x_{\rm
HI}$=0.971 (0.974), 0.519 (0.494) and 0.148 (0.151) at $z$=14, 9 and
7, respectively.  Despite the fact that such a small box is not able
to capture the reionization process as a whole, it still matches the
available observables as the larger boxes do, and it is more
appropriate to investigate resolution effects.  Also in this case the
outputs of the simulations are sufficiently close to each other in
redshift space that there is always some overlap between subsequent
boxes. Differently from earlier though, we have now a larger number of
boxes $B_i$ (with $i=1...450$) spanning the same redshift range $z
\approx 6-15$.

\subsection{Optical depth to the 21~cm line}
\label{sec:tau21cm}

In the upper panel of Figure~\ref{fig:tau21cm} we show the evolution
of the volume-averaged (over the entire box) 21~cm optical depth for
all models.  The value of $\tau_{\rm 21cm}$ is the same in all models
at the highest redshifts, when the evolution of both the HI fraction
and temperature are similar (see C2012 for more details). Differences
are more evident at lower $z$. In particular, in model ${\mathcal
E}$1.2-$\alpha$3 $\tau_{\rm 21cm}$ is up to one order of magnitude
higher, because reionization is delayed compared to the other cases,
for which differences are smaller. More specifically,
${\mathcal E}$1.6-$\alpha$3 has a similar $x_{\rm HI}$, but lower
temperature compared to ${\mathcal E}$1.2-$\alpha$1.8 and ${\mathcal
E}$1.2-$\alpha$1-3, resulting in a slightly higher value of the
optical depth. On the other hand, these latter two models have very
similar values for both $x_{\rm HI}$ and $T$, and the differences in
the volume averaged optical depth arise from more subtle details of
the distribution of the physical quantities.  As mentioned above, the
reionization history in run ${\mathcal L}$4.39 is very similar to that
of ${\mathcal E}$1.2-$\alpha$3, with a slightly (few percent) lower value 
of $x_{\rm HI}$ and a slightly (few percent) larger value of
the volume averaged temperature\footnote{Despite having the same mean
density of the larger box, the 4.39$h^{-1}$ box has gas at higher
density. Because of this, more
ionizing photons are needed to maintain a similar ionisation fraction,
with the consequence of an increased photo-heating.}. The above
results in a lower $\tau_{\rm 21cm}$.

\begin{figure}
\centering
\includegraphics[width=0.45\textwidth]{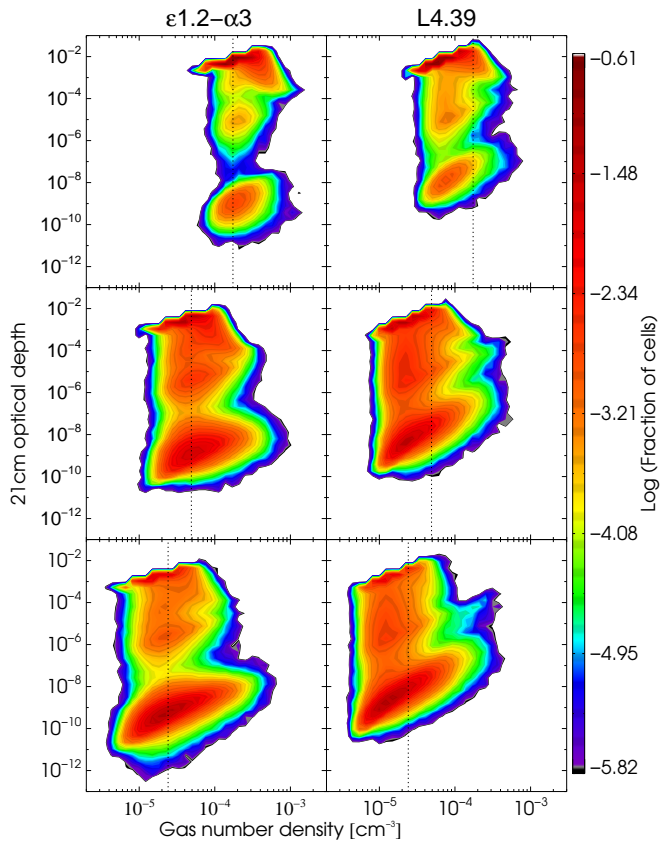}
\caption{Distribution of the 21~cm optical depth versus gas number
density for model ${\mathcal E}$1.2-$\alpha$3 (left panel) and
${\mathcal L}$4.39 (right).  The rows refer to redshift $z=14, 9$ and
7 (from top to bottom). The vertical lines refer to the
volume-averaged gas number density in the box, which corresponds to
$\delta=1$.  }
\label{fig:contours_tau21cm}
\end{figure}

In the lower panel of Figure~\ref{fig:tau21cm} we show the evolution of the
volume averaged 21~cm optical depth in model ${\mathcal
E}$1.2-$\alpha$3 for different assumptions about the spin temperature,
$T_s$.  The green dotted-dashed line is our reference case, in which
$T_s$ is assumed to be in equilibrium with the gas temperature, $T$,
calculated from the radiative transfer simulation (which includes only
UV photons). The same result is obtained when $T_s$ is calculated as
the maximum between the gas temperature as determined from the
radiative transfer calculation and that including also the
contribution from Ly$\alpha$ photons from stars. If in addition to the
Ly$\alpha$ photons we consider also the contribution from x-ray
photons from quasars (red dotted line), the gas temperature increases
and as a consequence the optical depth decreases.  The temperature
with Ly$\alpha$ and x-ray photons is calculated for a gas at the mean
density using the semi-analytic approach as described in
\cite{Ciardi.Salvaterra.DiMatteo_2009}.  Here, the authors calculate the photon 
production from a population of stars and micro-quasars in a set of cosmological 
hydrodynamic simulations \citep{Pelupessy.DiMatteo.Ciardi_2007}
which self-consistently follow the dark matter dynamics, 
radiative processes as well as star formation, black hole growth and associated 
feedback processes. We refer the reader to the original papers for more details.
More specifically, the gas
temperature as determined by Ly$\alpha$ photons from stars is given by
the short-dash-dotted blue line in Figure~2 of that paper, while the
one with also heating from x-rays from quasars corresponds to the
solid black line in the same plot. While in cells which have been
(also partially) ionized the temperature is $\sim 10^4$~K, cells which
have not been reached by ionizing photons have a temperature which can
be as cold as that of the CMB. In this case, a contribution to the
heating from Ly$\alpha$ or x-ray photons can raise the temperature
above its original value, reducing the optical depth to 21~cm.  It
should be noted that the above results depend on the model
adopted for the Ly$\alpha$ and x-ray photon production, and that this
is not self-consistent with the RT calculations. While we will address
this point more thoroughly in future developments by running radiative transfer 
simulations
which follow self-consistently also the propagation of x-ray and Ly$\alpha$
photons, the qualitative results will still hold.

As a reference, the $\tau_{\rm 21cm}$ calculated in model ${\mathcal
E}$1.2-$\alpha$3 is $\sim 16\%$ (50\%) lower than the one found by
\cite{Xu_etal_2009} at $z=13.5$ (9.3), and it becomes $\sim 100 \%$
larger at $z=6$.  These values are lower than those
found by \cite{Mack.Wyithe_2011} in their model without x-ray heating,
most probably because they obtain an average gas temperature which is
never above $\sim 1000$~K (also when x-ray heating is included) in the
relevant redshift range, i.e. about one order of magnitude lower than
the one found in our models.

It should be noted that in all cases $\tau_{\rm 21cm}$ is just an
average value, but the scatter is very large, as can be seen in
Figure~\ref{fig:contours_tau21cm}, where the distribution of 21~cm
optical depth versus gas number density is shown for two reference
simulations at redshifts $z=14$ (top panels), 9 (middle panels) and 7
(lower panels).  In the 35$h^{-1}$~Mpc box, most of the gas is
initially concentrated either in high density and high optical depth
cells (corresponding to cold, neutral, high density regions) or in
cells with $\tau_{\rm 21cm}\sim 10^{-9}$ (corresponding to gas which
has been ionized and heated to high temperatures).  As reionization
proceeds and more cells get ionized, the corresponding optical depth
decreases. We also observe a larger spread of $\tau_{\rm 21cm}$ with
the gas number density, due to the fact that photons have now reached
low density regions far away from the location of the sources of
radiation. While at $z=9$ there is still a large range of values of
the optical depth, at $z=7$ most cells have $\tau_{\rm 21cm} \sim
10^{-10}$, because reionization is almost complete.  It is interesting
to compare the distribution of the optical depth to the one of the gas
temperature shown in Figure~7 of C2012. In fact, the two quantities
exhibit a complementary behaviour, as expected.  The distribution and
evolution of $\tau_{\rm 21cm}$ for ${\mathcal L}$4.39 is similar,
although the optical depth is always higher, and the range of
densities spanned is larger compared to the other models.

\begin{figure}
\centering
\includegraphics[width=0.40\textwidth]{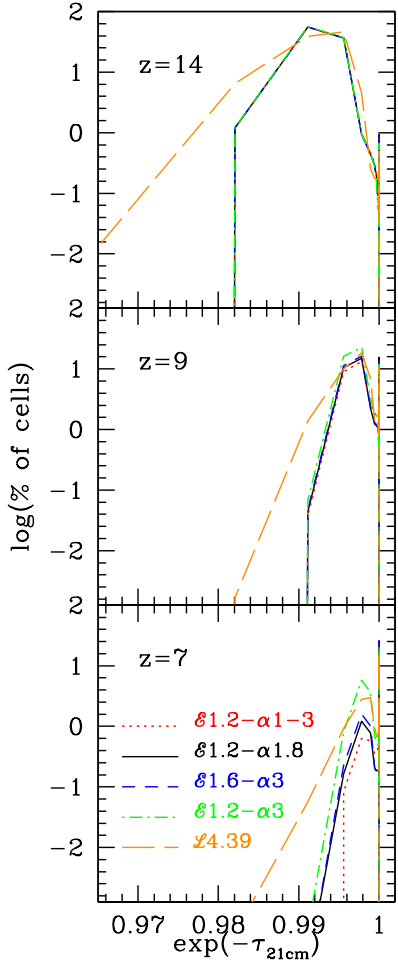}
\caption{Percentage of cells as a function of the transmissivity,
  exp(-$\tau_{\rm 21cm}$), at $z$=14 (upper panel), 9 (middle) and 7
  (lower). The black solid, red dotted, blue dashed, green
  dotted-dashed and orange long-dashed lines indicate results for
  models ${\mathcal E}$1.2-$\alpha$1.8, ${\mathcal E}$1.2-$\alpha$1-3,
  ${\mathcal E}$1.6-$\alpha$3, ${\mathcal E}$1.2-$\alpha$3 and
  ${\mathcal L}$4.39, respectively.  }
\label{fig:distr_trans}
\end{figure}

A more quantitative representation is given in
Figure~\ref{fig:distr_trans}, where we show the percentage of cells as
a function of the transmissivity, exp(-$\tau_{\rm 21cm}$), i.e.  the
quantity most relevant to observations (see following Sec.).  In all
models, at the highest redshift most of the cells have a high optical
depth (i.e. exp(-$\tau_{\rm 21cm})<1$), although, as already
mentioned, there is a large spread and the cells which have already
been ionized are concentrated around $\tau_{\rm 21cm} \sim 10^{-9}$
(i.e. exp(-$\tau_{\rm 21cm}) \sim 1$).  As ionization proceeds, the
number of cells with such optical depth increases, while those with
higher $\tau_{\rm 21cm}$ decreases. This results in a shift towards
larger values of the transmissivity.  At all redshifts there are also
intermediate values of $\tau_{\rm 21cm}$, reflecting the distribution
of densities and temperatures (see C2012).  The limitations due to resolution
effects are visible at all redshifts as a sudden drop of the number of
cells below a minimum transmissivity.  The behaviour is very similar
in all models, with a systematically lower transmissivity for
${\mathcal L}$4.39.  Due to the higher resolution of ${\mathcal
L}$4.39, larger values for the gas overdensity are
sampled in this simulation.  It should be noted though that also in
this smaller simulation the resolution is not enough to capture the
even higher overdensities which are associated with collapsed objects.
This topic is the focus of a future investigation.

\begin{figure}
\centering
\includegraphics[width=0.40\textwidth]{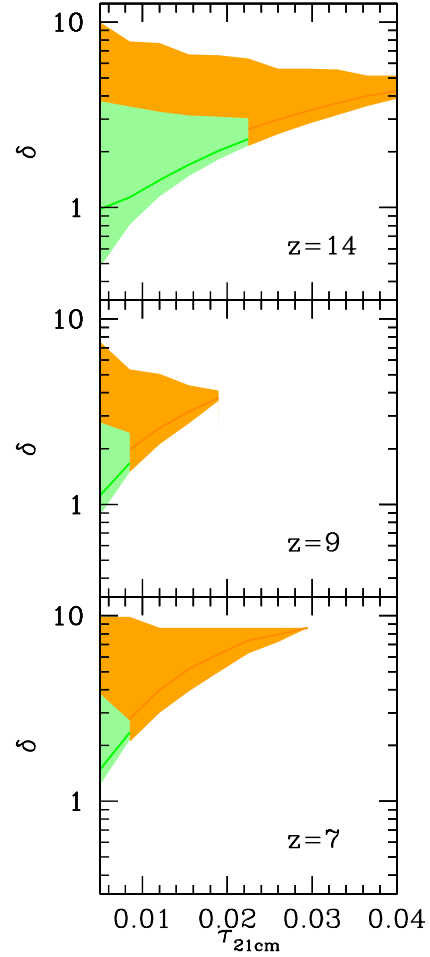}
\caption{Gas overdensity of cells with optical depth larger than
  $\tau_{\rm 21cm}$ at redshift $z=14$ (upper panel), 9 (middle) and 7
  (lower) for models ${\mathcal L}$4.39 (orange shaded area in the
  back) and ${\mathcal E}$1.2-$\alpha$3 (green shaded area in the
  front). The shaded areas refer to the full range of overdensities, 
  while the solid lines indicate the mean  values. }
\label{fig:delta}          
\end{figure}

In Figure~\ref{fig:delta}
the gas overdensity of cells with optical depth larger than $\tau_{\rm 21cm}$ is 
plotted at redshift $z=14$, 9  and 7 for models
${\mathcal L}$4.39 and ${\mathcal E}$1.2-$\alpha$3.
The shaded areas refer to the full range of
overdensities, while the solid lines indicate the mean values. It is evident that,
while the mean value is similar in both cases (because the hydrodynamical simulations
have the same mean density), ${\mathcal L}$4.39 reaches higher overdensities which
produce the larger values of the optical depth. 
This is important for the observability of the line, as we will see in the following.

\section{Mock spectra}
\label{sec:spectra}

\begin{figure}
\centering
\centering
\includegraphics[width=0.5\textwidth]{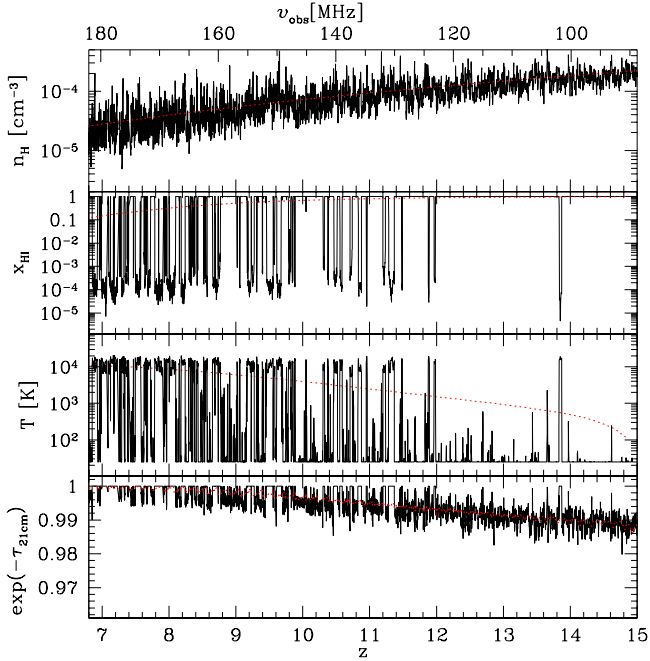}
\caption{Physical state of the gas along a random LOS (solid lines) in
  model ${\mathcal E}$1.2-$\alpha$3.  From top to bottom: physical H
  number density, \HI fraction, gas temperature, gas
  transmissivity. The red dotted lines indicate the volume-averaged
  values of the different quantities.}
\label{fig:los}
\end{figure}

\begin{figure}
\centering
\centering
\includegraphics[width=0.5\textwidth]{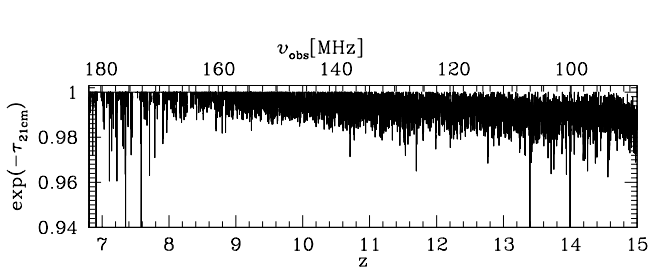}
\caption{Gas transmissivity along a random LOS in model ${\mathcal
L}$4.39.  }
\label{fig:los_Lbox4.39}
\end{figure}

From each simulation box $B_i$ we extract 2500 random LOS. A total LOS
is obtained as follows.  We choose randomly a LOS ($los_1$) among
those built from box $B_1$ and we stack it with one ($los_2$) chosen
randomly among those built from box $B_2$. Whenever $los_1$ and
$los_2$ overlap in redshift, we average the relevant quantities. For
example, $n_{\rm HI}(z)$ of the total LOS is given by $[n_{\rm
HI,i}(z)+n_{\rm HI,i+1}(z)]/2$, where $n_{\rm HI,i}(z)$ is the \HI
number density at redshift $z$ in $los_i$.  This procedure is repeated
for all the boxes.  

An example of such a LOS from model ${\mathcal
E}$1.2-$\alpha$3 is shown in Figure~\ref{fig:los}. Here the panels
(from top to bottom) indicate the evolution of H number density, \HI
fraction, gas temperature and transmissivity, $e^{-\tau_{\rm 21cm}}$.
The dotted lines indicate the volume-averaged values of the different
quantities. It should be noted that here no smoothing has been applied
and the resolution shown corresponds to the one of the simulation,
i.e. $\sim$15-20~kHz depending on redshift. We will discuss the higher
resolution simulations in Section~\ref{sec:instrument}, while here
(Fig.~\ref{fig:los_Lbox4.39}) we just show the transmissivity along a
random LOS as a comparison with the larger box.  Below, we will
discuss the dependence of our results on a number of quantities, using
the simulations run in the 35$h^{-1}$~Mpc box.

\subsection{Dependence on spin temperature}
\label{sec:spin}

\begin{figure}
\centering
\includegraphics[width=0.5\textwidth]{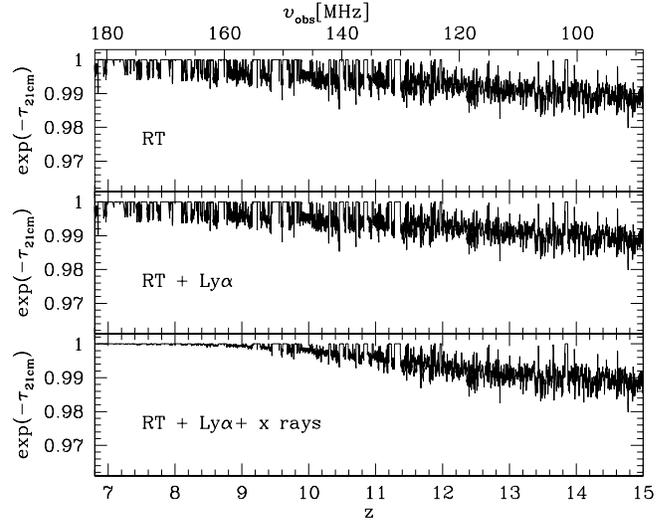}
\caption{Evolution of the gas transmissivity along a random LOS for
  different assumptions about the gas temperature, $T$.  From top to
  bottom the gas temperature is: the one determined from the radiative
  transfer of UV photons only; the maximum of the one determined from
  the radiative transfer of UV photons and the one calculated
  including also the contribution from Ly$\alpha$ photons from stars; the
  maximum of the one determined from the radiative transfer of UV
  photons and the one calculated including also the contribution from
  Ly$\alpha$ photons from stars and x-ray photons from quasars.  }
\label{fig:spectra.ts}
\end{figure}

In Figure~\ref{fig:spectra.ts} the dependence on different assumptions
about the gas spin temperature is shown.  The LOS is the same as that
plotted in Figure~\ref{fig:los}, but now we show not only the reference
case (upper panel), but also a case in which $T_s$ is taken to be the
maximum of the gas temperature as determined from the radiative
transfer calculation (which includes only UV photons) and that
including also the contribution from Ly$\alpha$ photons from stars
(middle panel), and from Ly$\alpha$ plus x-ray photons from quasars
(lower panel). More details on the calculation of the temperature are
given in Section~\ref{sec:tau21cm}.  Because Ly$\alpha$ heating is not
extremely efficient, the resulting transmissivity is very similar to the
case without such heating. On the contrary, the contribution from
x-rays is more substantial and this can clearly be seen in the Figure
as a reduction of the transmissivity at redshifts when the
extra heating is significant ($z \simlt 12$).  This suggests that in
the presence of x-ray heating, the detection of the 21~cm forest would most
probably not be feasible (but see e.g. \citealt{Mack.Wyithe_2011} and
\citealt{Xu.Ferrara.Chen_2011} for a discussion of the dependence of
absorption features on the strength of the x-ray heating).
It should be noted that the value of the x-ray background
used here is appropriate for a population of high-$z$ quasars. As a reference,
at $z=10$ it is $\sim 4 \times 10^{-26}$~ergs~cm$^{-2}$~s$^{-1}$~Hz$^{-1}$~sr$^{-1}$
(see Fig.~1 of \citealt{Ciardi.Salvaterra.DiMatteo_2009}).
If the background were instead dominated by a different class of sources, as
x-ray binaries (see e.g. \citealt{Furlanetto_2006,Dijkstra.Gilfanov.Loeb.Sunyaev_2012}),
then the associated heating would be different. We defer to future work a
more detailed investigation of the role of the x-ray background and the contribution to it
from different types of sources.

As already pointed out by other authors
(e.g. \citealt{Mack.Wyithe_2011}), the above arguments seem to suggest
that it would be possible to discriminate between different IGM
reheating histories, in particular if a high energy component 
were present in the ionising spectrum.

\subsection{Dependence on source spectra}
\label{sec:sources}

\begin{figure}
\centering
\includegraphics[width=0.5\textwidth]{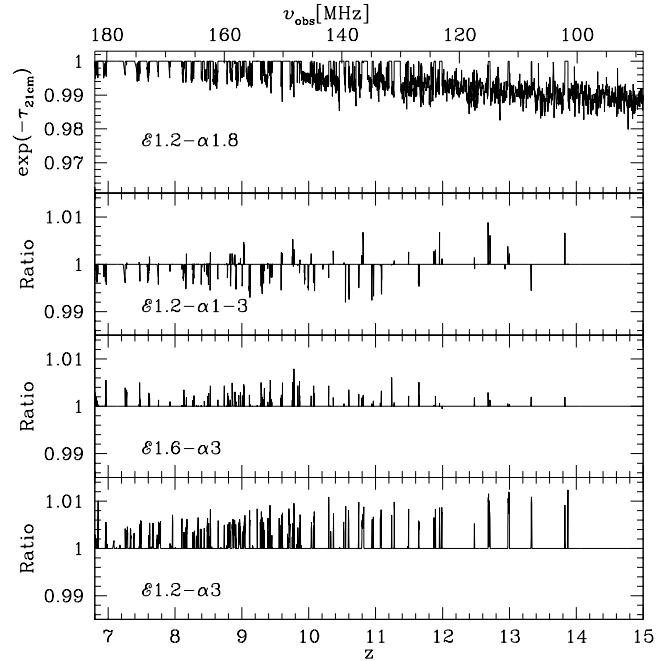}
\caption{From the top to the bottom panels we show: the evolution of
  the gas transmissivity along a random LOS for reionization model
  ${\mathcal E}$1.2-$\alpha$1.8; the ratios between the transmissivity
  of model ${\mathcal E}$1.2-$\alpha$1.8 and models ${\mathcal
  E}$1.2-$\alpha$1-3, ${\mathcal E}$1.6-$\alpha$3, and ${\mathcal
  E}$1.2-$\alpha$3, respectively.  The ratios have been calculated using
  the same LOS as the upper panel.  }
\label{fig:spectra.spectra}
\end{figure}

As discussed in Section~\ref{sec:simul}, the simulations have been run
for a variety of source spectra.  Here we show the differences induced
in terms of transmission.  In the top panel of
Figure~\ref{fig:spectra.spectra} a reference LOS for reionization
model ${\mathcal E}$1.2-$\alpha$1.8 is shown.  In the other panels the
ratios between the transmissivity of model ${\mathcal
E}$1.2-$\alpha$1.8 and models ${\mathcal E}$1.2-$\alpha$1-3,
${\mathcal E}$1.6-$\alpha$3, and ${\mathcal E}$1.2-$\alpha$3 are
plotted. In the third and fourth panels from the top the ratio is
always larger than 1, meaning that the optical depth in ${\mathcal
E}$1.2-$\alpha$1.8 is lower than in ${\mathcal E}$1.6-$\alpha$3 and
${\mathcal E}$1.2-$\alpha$3 (i.e. the simulations with softer
spectra), as expected.  In fact, although $x_{\rm HI}$ is very similar
in ${\mathcal E}$1.2-$\alpha$1.8 and ${\mathcal E}$1.6-$\alpha$3, the
temperature in the latter is typically lower, resulting in a higher
optical depth. On the other hand, as in model ${\mathcal
E}$1.2-$\alpha$3 reionization is less advanced, the optical depth is
always larger than for the other cases.  Models ${\mathcal
E}$1.2-$\alpha$1.8 and ${\mathcal E}$1.2-$\alpha$1-3 have very similar
average $x_{\rm HI}$ and $T$, but their distributions are slightly
different. This reflects also on the transmissivity.

These very small differences in transmissivity predicted by different
reionization models are to be expected. As discussed in the previous
Section, the models do have different values of $\tau_{\rm 21cm}$, but
these are more pronounced when $\tau_{\rm 21cm}$ is in any case too
low to produce appreciable absorption.  In fact, all models have
very similar hydrogen reionization histories, with the largest
discrepancies being associated with the temperature reached either in
the ionized regions or in their outskirts, where the hardness of the
spectrum makes most of the difference.  From the point of view of the
absorption, though, the exact value of the temperature in a fully or
partially ionized cell is not relevant. On the other hand, the
regions which produce most of the absorption, i.e. the cold, high
density gas not reached by ionizing photons, are basically the same in
all models. Thus, unless the reionization histories are very different
from each other (e.g. for example a predominance of UV vs. x-ray
sources), we do not expect to be able to distinguish them by means of observations
of the 21~cm forest.


\section{Instrumental effects}
\label{sec:instrument}

In this Section we describe the process used to simulate the spectra
as they would be observed by {\tt LOFAR}.  The input of the simulation
is a file containing the optical depth as a function of redshift (or,
equivalently, frequency) and the parameters associated with the
background radio source, i.e. its redshift, $z_s$, flux density,
$S_{\rm in}(z_s)$, and spectral index, $\alpha$. Consistently with
previous studies (e.g. \citealt{Carilli.Gnedin.Owen_2002}) we assume
that the source has a power-law spectrum.  In principle, the effect of
the Point Spread Function (PSF) side lobes running through the source
of interest can be taken into account by including more sources at
different positions, with or without absorption features. Here, this
has not been considered.  For a narrow bandwidth, the equation giving
the observed visibilities is:

\begin{equation}
{V_\nu }\left( {\bf{u}} \right) = \sum\limits_i^{{N_{sources}}} {{I_\nu }({\bf{s}})} 
{e^{ - 2\pi i{\bf{u}} \cdot {\bf{s}}}} + n,
\end{equation}
where $\mathbf{u}=(u, v, w)$ are the  coordinates of a given baseline at a certain 
time $t$, ${I_\nu }$ is the observed source intensity, $\mathbf{s}=(l,m, n)$ is 
a vector representing the direction cosines for a given source direction and $n$ 
represents additive noise.  The noise is given by the radiometer equation:

\begin{equation}
{\sigma _n} = \frac{1}{{{n_s}}}\frac{{SEFD}}{{\sqrt {N\left( {N - 1} \right)
{t_{{\mathop{\rm int}} }} \Delta \nu} }},
\label{eq:noise}
\end{equation}
where $n_s$ is the system efficiency, $\Delta\nu$ is the bandwidth, $t_{int}$ is 
the integration time and $N$ is the number of stations. We assume that  $n_s=0.5$
and $N=48$.  The system equivalent density is given by:

\begin{equation}
SEFD = \frac{{2{\kappa _B}{T_{\rm sys}}}}{{N_{\rm dip}{\eta _\alpha }{A_{\rm eff}}}},
\end{equation}
where $\kappa _B$ is Boltzmann's constant, $A_{\rm
eff}=min(\frac{\lambda2}{3}, 1.5626)$ is the effective area of each
dipole in the dense and sparse array regimes respectively, $N_{\rm
dip}$ is the number of dipoles per station (24 tiles times 16 dipoles
per tile for a {\tt LOFAR} core station) and $\eta_{\alpha}$ is the
dipole efficiency which we assume to be 1. The system noise $T_{\rm
sys}$ has two contributions: {\it (i)} from the electronics and {\it
(ii)} from the sky. We assume that the sky has a spectral index of
-2.55, obtaining $T_{\rm sys}=[140+60(\nu/150 \; {\rm MHz})^{-2.55}]$~K.

The complete Fourier plane sampling can be done by evaluating the
above equation for every set of baseline coordinates. The predicted
visibilities are then gridded and transformed via inverse Fourier
transforms in order to obtain the dirty images.

For the purpose of the 21~cm forest, fine spectral resolution is
needed, which means that the spectra need to be predicted for a large
number of channels ($\sim 10\, 000$).  Thankfully, the computation can
be done independently for each channel, and the maps can thus be
evaluated in parallel.  After assembling the full image cube, the LOS
spectrum is extracted.

\begin{figure*}
\centering
\includegraphics[width=0.7\textwidth]{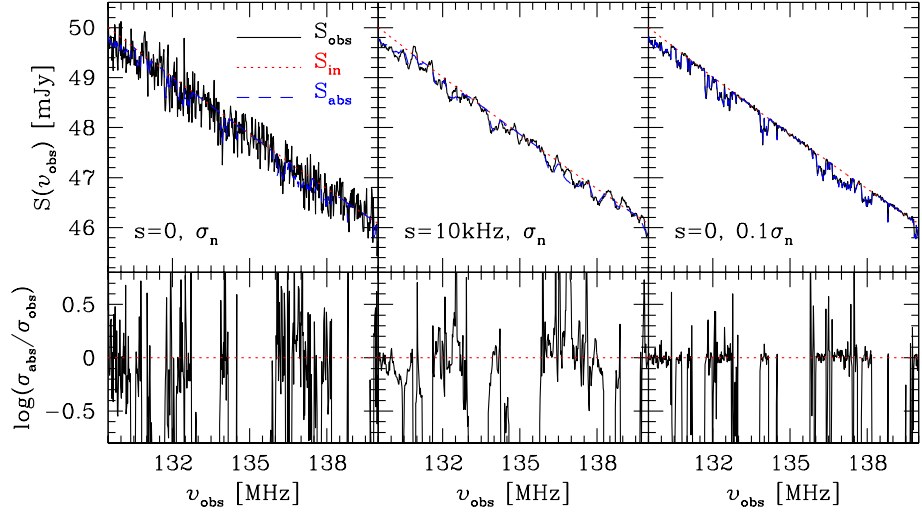}
\caption{{\it Upper panels:} Spectrum of a source positioned at
   $z_s=10$ (i.e. $\nu \sim 129$~MHz), with a power-law index of
   $\alpha=1.05$ and a flux density $S_{\rm in}(z_s)=50$~mJy.  The red
   dotted lines refer to the intrinsic spectrum of the radio source,
   $S_{\rm in}$; the blue dashed lines to the simulated spectrum for
   21~cm absorption, $S_{\rm abs}$; and the black solid lines to the
   spectrum for 21~cm absorption as it would be seen by {\tt
   LOFAR}. The latter has been calculated assuming an observation time
   $t_{int}=1000$~h and a bandwidth $\Delta \nu=20$~kHz. The IGM
   absorption is calculated from the reference simulation ${\mathcal
   E}$1.2-$\alpha$3.  Left panel: $S_{\rm abs}$ and $S_{\rm obs}$ are
   obtained without smoothing, $s=0$, and with the noise $\sigma_n$
   given in eq.~\ref{eq:noise}.  Middle panel: $S_{\rm abs}$ and
   $S_{\rm obs}$ are obtained after smoothing over a scale $s=10$~kHz, and
   with the noise $\sigma_n$ given in eq.~\ref{eq:noise}.  Right
   panel: $S_{\rm abs}$ and $S_{\rm obs}$ are obtained without
   smoothing, $s=0$, and with 1/10th of the noise $\sigma_n$ given in
   eq.~\ref{eq:noise}.  {\it Lower panels:} $\sigma_{\rm
   abs}/\sigma_{\rm obs}$ corresponding to the upper panels. Here $\sigma_i=
   S_i-S_{\rm in}$, with $i=$ abs, obs. See text for further details.  }
\label{fig:spectrum_z10_J50_Lbox35}
\end{figure*}

In the upper panels of Figure~\ref{fig:spectrum_z10_J50_Lbox35} we
show the spectrum of a source positioned at $z_s=10$ (i.e. $\nu \sim
129$~MHz). The intrinsic radio source spectrum, $S_{\rm in}$, is
assumed to be similar to Cygnus A (see also
\citealt{Carilli.Gnedin.Owen_2002,Mack.Wyithe_2011}), with a power-law
index $\alpha=1.05$ and a flux density $S_{\rm in}(z_s)=50$~mJy. 
It should be noted that if the intrinsic radio spectrum were not
a pure power-law, the identification of absorption features would be
rendered even more complex by the requirement that the continuum level
must be fit as well.
The simulated absorption spectrum, $S_{\rm abs}$, is calculated from the
IGM optical depth found in our reference simulation ${\mathcal
E}$1.2-$\alpha$3. The observed spectrum, $S_{\rm obs}$, is calculated
assuming an observation time $t_{int}=1000$~h and a bandwidth $\Delta
\nu=20$~kHz.  The spectra are plotted at the resolution of the
simulation, unless stated otherwise.  While some absorption features
are evident, it is not obvious that we will be able to disentangle
them from the instrumental noise. Even though our simulation box reaching
gas overdensities $\delta$ as high as a few tens, typically a LOS
through the IGM samples mild overdensities $\delta \sim 1$.  As an
example, while $\sim$40\% of cells along the LOS in the Figure (in the
full redshift range $z=6.5-15$) have $\tau_{\rm 21cm}>0.005$, none has
$\tau_{\rm 21cm}>0.025$, setting a limit on the gas transmissivity of
e$^{-\tau_{\rm 21cm}}>0.975$. Thus, similarly to
e.g. \cite{Mack.Wyithe_2011}, we do not find narrow strong absorption
lines (see below for further discussion).  If we do a Gaussian
smoothing of the spectrum over a scale $s=10$~kHz (upper middle panel)
or reduce the noise by a factor of 0.1 (similar to what is expected
from the {\tt SKA}; upper right panel), then the observability improves.
This is more evident in the lower panels of the Figure, which show the
quantity $\sigma_{\rm abs}/\sigma_{\rm obs}$, where
$\sigma_i=S_i-S_{\rm in}$ and $i$=abs, obs.

\begin{figure*}
\centering
\includegraphics[width=0.95\textwidth]{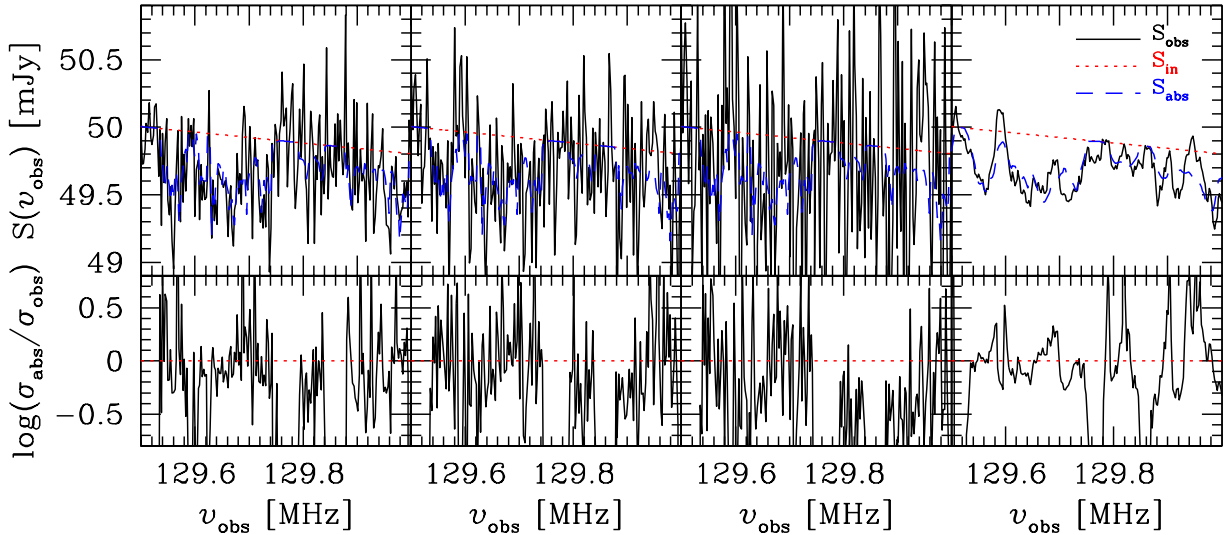}
\caption{{\it Upper panels:} Spectrum of a source positioned at $z=10$
  (i.e. $\nu \sim 129$~MHz), with an index of the power-law
  $\alpha=1.05$ and a flux density $S_{\rm in}(z_s)=50$~mJy.  The
  lines are the same as those in
  Figure~\ref{fig:spectrum_z10_J50_Lbox35}.  $S_{\rm obs}$ has been
  calculated assuming the noise $\sigma_n$ given in
  eq.~\ref{eq:noise}.  The IGM absorption is calculated from the
  reference simulation ${\mathcal L}4.39$.  From left to right the
  panels refer to a case in which $S_{\rm abs}$ and $S_{\rm obs}$ are
  obtained: with a bandwidth $\Delta \nu=20$~kHz, without smoothing,
  $s=0$, and with an integration time $t_{int}=1000$~h; with a
  bandwidth $\Delta \nu=20$~kHz, without smoothing, $s=0$, and with an
  integration time $t_{int}=500$~h; with a bandwidth $\Delta
  \nu=2.5$~kHz, smoothing over a scale $s=2.5$~kHz, and with an
  integration time $t_{int}=1000$~h; with a bandwidth $\Delta
  \nu=10$~kHz, smoothing over a scale $s=10$~kHz, and with an
  integration time $t_{int}=1000$~h.  {\it Lower panels:} $\sigma_{\rm
  abs}/\sigma_{\rm obs}$ corresponding to the upper panels.  }
\label{fig:spectrum_z10_J50_Lbox4.39}
\end{figure*}

Stronger absorption features appear if we use model ${\mathcal
L}4.39$, in which larger overdensities are present (see
Section~\ref{sec:tau21cm}). This is visible in
Figure~\ref{fig:spectrum_z10_J50_Lbox4.39}, where $S_{\rm in}$,
$S_{\rm abs}$ and $S_{\rm obs}$ are shown for the same source of
Figure~\ref{fig:spectrum_z10_J50_Lbox35}. The frequency range plotted
is smaller to allow a better visualization of the absorption features.
In this case the prospect for a detection improves, even halving the
observation time (or, equivalently, the bandwidth). The third (fourth)
panel from the left shows the case for $\Delta \nu=2.5$~kHz ($\Delta
\nu=10$~kHz) and a smoothing of the spectrum over the same
scale. While retaining a high resolution might result in a null
detection, observation features are clearly visible if a larger
bandwidth and smoothing scale are used.  We have verified that
features are detected for $\Delta \nu \simgt 5$~kHz.  Similarly to
model ${\mathcal E}$1.2-$\alpha$3, along a LOS in the same redshift
range, here we have $\sim$40\% of cells with $\tau_{\rm 21cm}>0.005$,
but we also find a handful of cells with $\tau_{\rm 21cm}>0.1$,
corresponding to a gas transmissivity e$^{-\tau_{\rm 21cm}}<0.905$.

\begin{figure}
\centering
\includegraphics[width=0.50\textwidth]{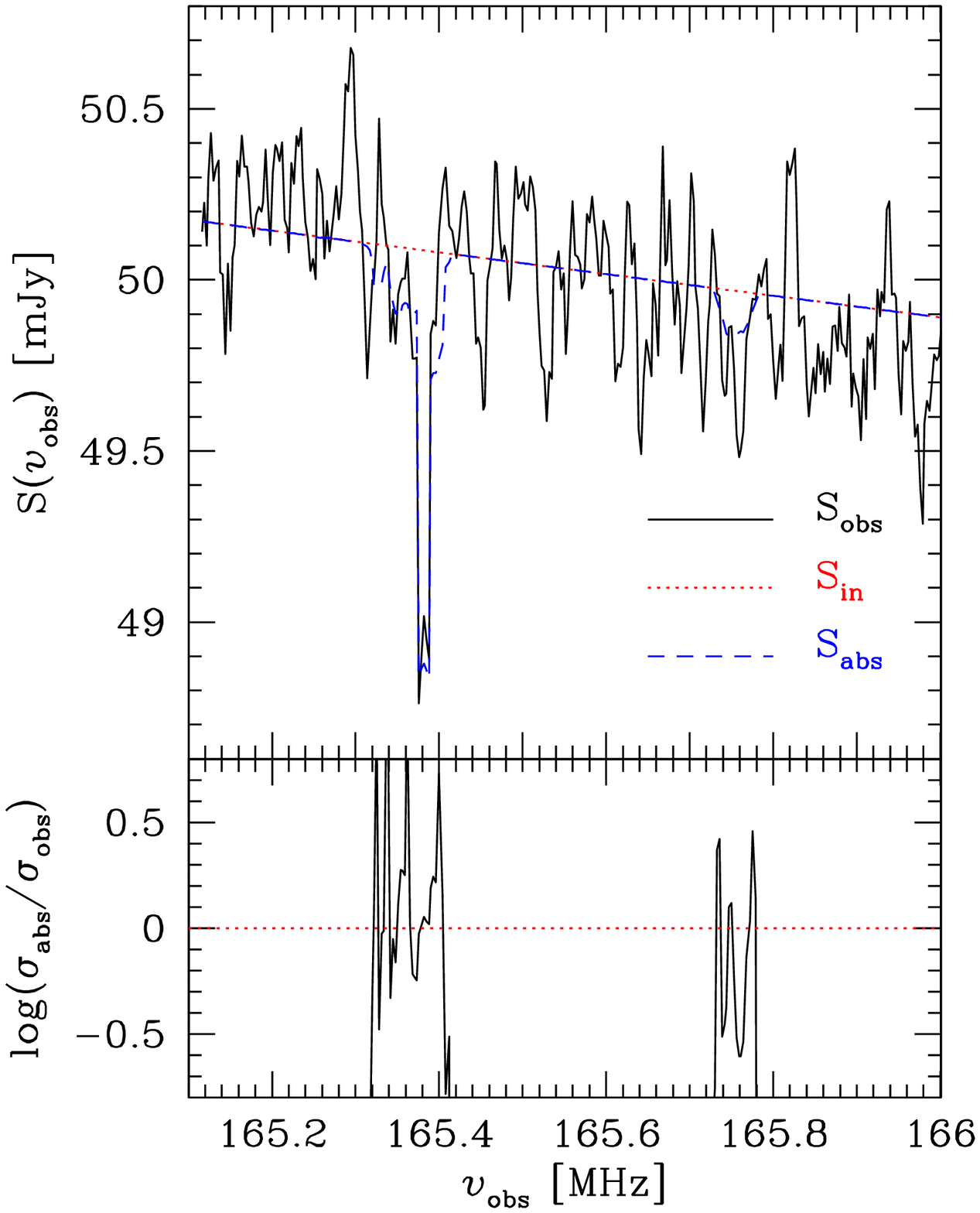}
\caption{{\it Upper panel:} Spectrum of a source positioned at $z=7.6$
  (i.e. $\nu \sim 165$~MHz), with an index of the power-law
  $\alpha=1.05$ and a flux density $S_{\rm in}(z_s)=50$~mJy.  The
  lines are the same as those in
  Figure~\ref{fig:spectrum_z10_J50_Lbox35}. Here we have assumed the
  noise $\sigma_n$ given in eq.~\ref{eq:noise}, a bandwidth $\Delta
  \nu=5$~kHz, smoothing over a scale $s=5$~kHz, and an integration
  time $t_{int}=1000$~h.  The IGM absorption is calculated from the
  reference simulation ${\mathcal L}4.39$.  {\it Lower panel:}
  $\sigma_{\rm abs}/\sigma_{\rm obs}$ corresponding to the upper
  panel. 
  Note that this LOS has been chosen to intercept a strong absorption
  feature, with $\tau_{\rm 21cm}=0.12$.}
\label{fig:spectrum_z7.6_J50_Lbox4.39}
\end{figure}

If we were lucky enough to intercept one of such high density pockets
of gas  (with $\tau_{\rm 21cm}>0.1$; these cells are found in
$\sim$ 0.1\% of the LOS), very strong absorption features
could then be detected, as
shown in Figure~\ref{fig:spectrum_z7.6_J50_Lbox4.39}, where we have
located at $z=7.6$ a radio source with the same characteristics as
above. To calculate the observed spectrum we have now used a $\Delta
\nu=5$~kHz and an integration time $t_{int}=1000$~h.  An even better
prospect of detection would arise if e.g. a high redshift Damped
Ly$\alpha$ System (DLA) were intercepted, but the simulations employed
in this work do not have the resolution to investigate this
possibility, which will be addressed elsewhere.

\begin{figure}
\centering
\includegraphics[width=0.50\textwidth]{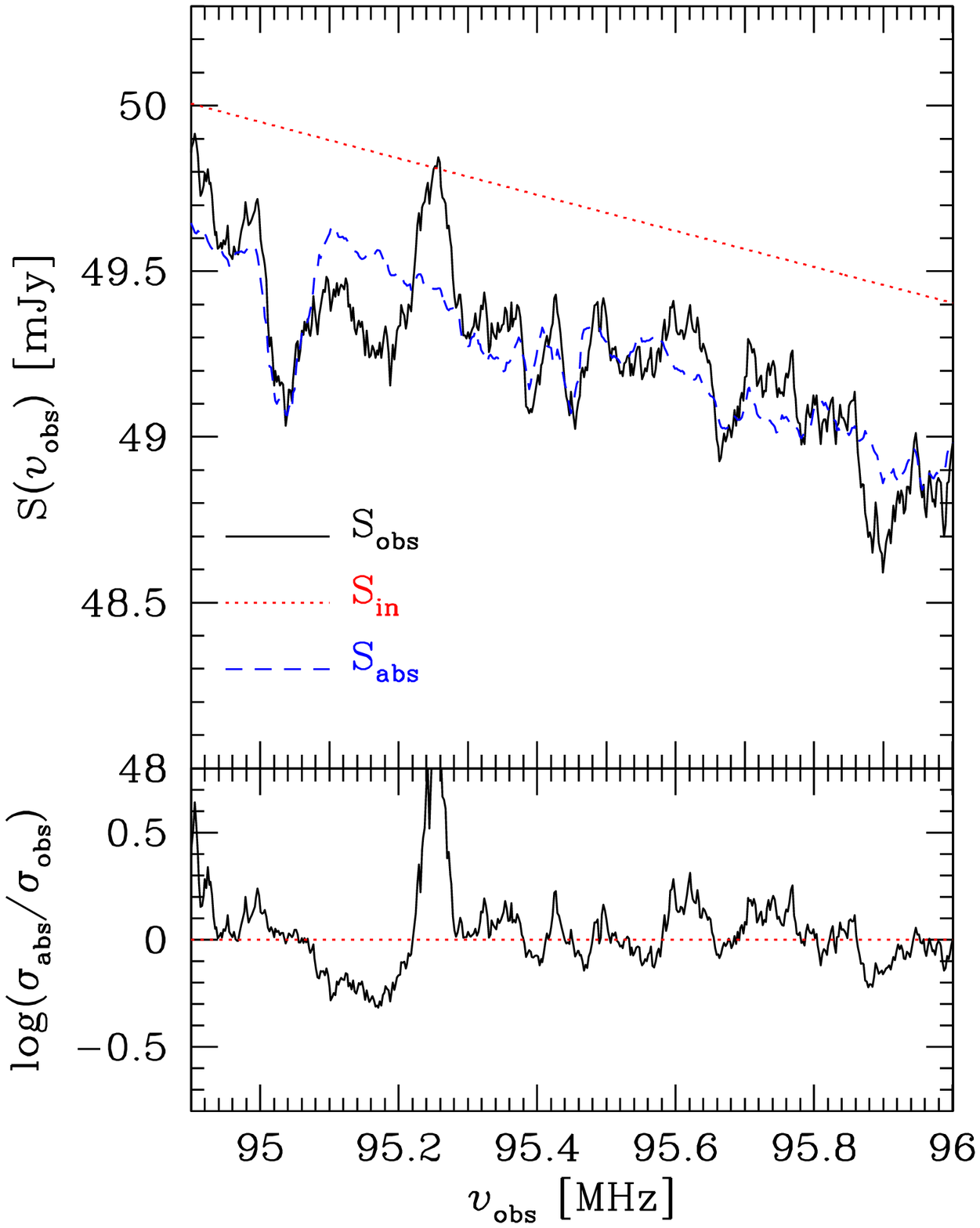}
\caption{{\it Upper panel:} Spectrum of a source positioned at $z=14$
  (i.e. $\nu \sim 95$~MHz), with an index of the power-law
  $\alpha=1.05$ and a flux density $S_{\rm in}(z_s)=50$~mJy.  The
  lines are the same as those in
  Figure~\ref{fig:spectrum_z10_J50_Lbox35}. Here we have assumed the
  noise $\sigma_n$ given in eq.~\ref{eq:noise}, a bandwidth $\Delta
  \nu=20$~kHz, smoothing over a scale $s=20$~kHz, and an integration
  time $t_{int}=1000$~h.  The IGM absorption is calculated from the
  reference simulation ${\mathcal L}4.39$.  {\it Lower panel:}
  $\sigma_{\rm abs}/\sigma_{\rm obs}$ corresponding to the upper
  panel.}
\label{fig:spectrum_z14_J50_Lbox4.39}
\end{figure}

Finally, moving towards higher redshift ($z > 12$), when most of the gas in the
IGM is still neutral and relatively cold, would offer the chance of
detecting a stronger average absorption.  Although this range of
redshift falls outside the one observed by {\tt LOFAR} because of
strong RFI contamination (Offringa et al. 2012, in prep), it is still
interesting to look at the example shown in
Figure~\ref{fig:spectrum_z14_J50_Lbox4.39}, where the usual reference
radio source has been located at $z=14$. The Figure clearly shows that
a {\tt LOFAR}-type telescope could easily detect the global absorption
(rather than the single absorption features observed at lower
redshift) due to the highly neutral IGM if a relatively large
bandwidth is used. In the Figure the results for $\Delta \nu =20$~kHz
are shown, but a suppression of the source flux could also be detected
for $\Delta \nu=10$~kHz.  It should be noted, however, that
distinguishing a suppression of the source flux due to intervening
neutral hydrogen from a source with an intrinsically lower flux would not
be straightforward.

\section{Conclusions}

In this paper we have discussed the feasibility of the detection of
the 21~cm forest in the diffuse IGM with the radio telescope {\tt
LOFAR}. The optical depth to the 21~cm line has been derived using the
simulations of reionization presented in
\citet{Ciardi.Bolton.Maselli.Graziani_2012}, which reproduce a number
of observational constraints such as the Thomson scattering optical
depth \citep{Komatsu_etal_2011} and the \HI photo-ionisation rate
at $z\sim 6$
\citep{Calverley.Becker.Haehnelt.Bolton_2011,Wyithe.Bolton_2011}. The
simulations provide the evolution of the spatial distribution of
relevant physical properties such as gas neutral fraction, temperature
and velocity field. The main results of our investigation can be
summarized as follows.
\begin{itemize}
\item The spectra from reionization models with similar total comoving hydrogen
  ionizing emissivities, but different frequency distributions look remarkably similar.
  Thus, unless the reionization histories are very different from each other
  (e.g. a predominance of UV vs. x-ray heating), we do not expect to distinguish them
  by means of observations of the 21~cm forest.
\item The photo-heating associated with the presence of a strong x-ray background would make the
  detection of the 21~cm line absorption impossible. The lack of
  absorption could then be used as a probe of the presence/intensity
  of the x-ray background and the thermal history of the universe.
\item Along a random line of sight {\tt LOFAR} could detect a global
  suppression of the spectrum from $z$ \simgt 12, when the IGM is
  still mostly neutral and cold, in contrast with the more well-defined, albeit
  broad, absorption features visible at lower redshift. Sharp, strong
  absorption features associated with rare, high density pockets of gas
  could be detected at $z \sim 7$ along preferential lines of
  sight.
\end{itemize}

The most challenging aspect of the detection of a 21~cm forest remains
the existence of high-$z$ radio loud sources. Although a QSO has been
detected at $z=7.085$ \citep{Mortlock_etal_2011}, the existence of
even higher redshift quasars is uncertain. The predicted number of
radio sources which can be used for 21~cm forest studies in the whole
sky per unit redshift at $z=10$ varies in the range $10-10^4$
depending on the model adopted for the luminosity function of such
sources and the instrumental characteristics
(e.g. \citealt{Carilli.Gnedin.Owen_2002, Xu_etal_2009}), making such a
detection an extremely challenging task. The possibility of using
GRB afterglows has been suggested by \cite{Ioka.Meszaros_2005},
concluding that it will be difficult to observe an absorption line,
even with the {\tt SKA}, except for very energetic sources, such as    
GRBs from the first stars. In
fact, a similar calculation has been repeated more recently by
\cite{Toma.Sakamoto.Meszaros_2011} for massive metal-free stars,
finding that the flux at the same frequencies should typically be at
least an order of magnitude higher than for a standard GRB.

An absorption feature stronger than the one produced by the diffuse
IGM, would be the one due to intervening starless minihalos or dwarf
galaxies (i.e. \citealt{Xu.Ferrara.Chen_2011,Meiksin_2011}), resulting
in an easier detection. On the other hand the optical depth would strongly
depend on the feedback effects acting on such objects. Because of the
large uncertainties about the nature and intensity of high-$z$ feedback
effects (for a review see \citealt{Ciardi.Ferrara_2005} and its ArXiv
updated version), it is not straightforward to estimate the relative
importance of these two absorption components unless a self-consistent
calculation is performed. We defer this investigation to a future
paper.

\section*{Acknowledgments}
The authors would like to thank an anonimous referee for his/her
comments.
This work was supported by DFG Priority Programs 1177 and 1573.  GH is
a member of the LUNAR consortium, which is funded by the NASA Lunar
Science Institute (via Cooperative Agreement NNA09DB30A) to
investigate concepts for astrophysical observatories on the Moon.
LVEK, HV and SD acknowledge the financial support from the European
Research Council under ERC-Starting Grant FIRSTLIGHT - 258942.

\bibliographystyle{apj} 
\bibliography{21cmForest.bib}

\label{lastpage}

\end{document}